\documentclass[aps, pre, twocolumn, superscriptaddress]{revtex4-1} 
\pdfoutput=1
\usepackage{amsmath}
\usepackage{amssymb}
\usepackage{graphicx}
\usepackage{url}
\usepackage[usenames,dvipsnames,svgnames,table]{xcolor}
\usepackage{hyperref}

\begin{document}

\title{Extracting hierarchical backbones from bipartite networks} 

\author{Woo Seong Jo}
\affiliation{The Center for Science of Science \& Innovation, Northwestern University, Evanston, IL, USA.}
\affiliation{Northwestern Institute on Complex Systems, Northwestern University, Evanston, IL, USA.}
\affiliation{Kellogg School of Management, Northwestern University, Evanston, IL, USA.}

\author{Jaehyuk Park}
\affiliation{Center for Complex Networks and Systems Research, Luddy School of Informatics, Computing, and Engineering, Indiana University, Bloomington, IN 47408, USA}

\author{Arthur Luhur}
\affiliation{Department of Biology, Indiana University, Bloomington, IN 47405, USA}

\author{Beom Jun Kim}\email{beomjun@skku.edu}
\affiliation{Department of Physics, Sungkyunkwan University, Suwon 16419, Republic of Korea}

\author{Yong-Yeol Ahn}\email{yyahn@iu.edu}
\affiliation{Center for Complex Networks and Systems Research, Luddy School of Informatics, Computing, and Engineering, Indiana University, Bloomington, IN 47408, USA}


\begin{abstract} 

We propose a method for extracting hierarchical backbones from a bipartite network. Our method leverages the observation that a hierarchical relationship between two nodes in a bipartite network is often manifested as an asymmetry in the conditional probability of observing the connections to them from the other node set. Our method estimates both the importance and direction of the hierarchical relationship between a pair of nodes, thereby providing a flexible way to identify the essential part of the networks. Using semi-synthetic benchmarks, we show that our method outperforms existing methods at identifying planted hierarchy while offering more flexibility. Application of our method to empirical datasets---a bipartite network of skills and individuals as well as the network between gene products and Gene Ontology (GO) terms---demonstrates the possibility of automatically extracting or augmenting ontology from data.

\end{abstract}

\maketitle

\section{Introduction}\label{sec:introduction}

As Herbert Simon argued~\cite{Simon1962}, the discovery of \emph{hierarchical structure} from networks---for instance in the form of community structure~\cite{Clauset2008, Blondel2008, yyAhn2010, Sales-Pardo-Amaral}, core-periphery structure~\cite{Hidalgo2007, Clauset_hiring}, or network backbones~\cite{Serrano2009, Radicchi2011, Grady2012}---has been one of the fundamental challenges in the study of networks and complex systems.
Although many approaches have been developed and applied to techno-social~\cite{Onnela2007, Newman2003, Sales-Pardo-Amaral, Balogh2019}, ecological~\cite{Dunne2002}, and biological systems~\cite{Goh2007}, most efforts are limited to the analysis of unipartite networks, where networks with only one node type is considered.

However, a large fraction of real-world networks are either bipartite or derived from underlying bipartite networks~\cite{Barabasi2016}.
For instance, unipartite social networks are commonly derived from bipartite networks of people and social groups~\cite{Barabasi1999, Amaral2000, Newman2001, Newman2002}; entity-tag networks are often projected to create similarity networks between entities~\cite{Zhou2007, Goh2007}.
Given that one-mode projection can destroy critical information in the original bipartite network~\cite{Lehmann2008, Larremore2014}, ability to fully leverage structure information in a bipartite network is critical for network science.

Here, we focus on the problem of extracting a `network backbone'---a sparse representation of the original network that captures the most important connections~\cite{Serrano2009, Radicchi2011, Grady2012}.
We extend the notion of the backbone and propose a method to extract a \emph{hierarchical} network backbone---which takes the form of DAG (Directed Acyclic Graph)---that captures connections that not only are structurally important, but also document strong hierarchical relationship.
In contrast to the most existing backbone-extraction methods, our approach leverages the underlying bipartite network structure to infer hierarchical relationships between one of the node sets.
We use an intuitive heuristic to estimate the direction and
strength of the hierarchical relationships between a pair of nodes in the network, addressing some of the weaknesses of the existing
approaches~\cite{Bishop_book, Heymann2006, Schmitz2006, Tibey2013}.

Let us explain the intuition behind our approach with an example.
Imagine a scientist, for whom we are picking disciplinary keywords from a hierarchically organized ontology.
If we pick ``Statistical Physics'' as their research area, it means that their research can also be described as ``Physics''.
One of the implications of this ontological relationship is that hierarchically related keywords are likely to co-occur.
Meanwhile, note that we can also expect, not only the co-occurrence, but also an asymmetry in the \emph{conditional probability} of observing these keywords.
That is, although ``Physics'' is a valid keyword for everyone with ``Statistical Physics'', the converse is not true because there are many subdisciplines (e.g. ``Optics'') in physics.
In other words, thanks to the ontological relationships between these keywords, for a pair of ontologically related keywords, we expect to see (i) significant co-occurrence as well as (ii) a strong asymmetry in the conditional probabilities of the co-occurrences.
Our approach leverages these two observations to identify hierarchical relationships and quantify their strengths.

\section{Results and Discussion}

\subsection{Hierarchical backbones}

\begin{figure}
\includegraphics[width=\columnwidth]{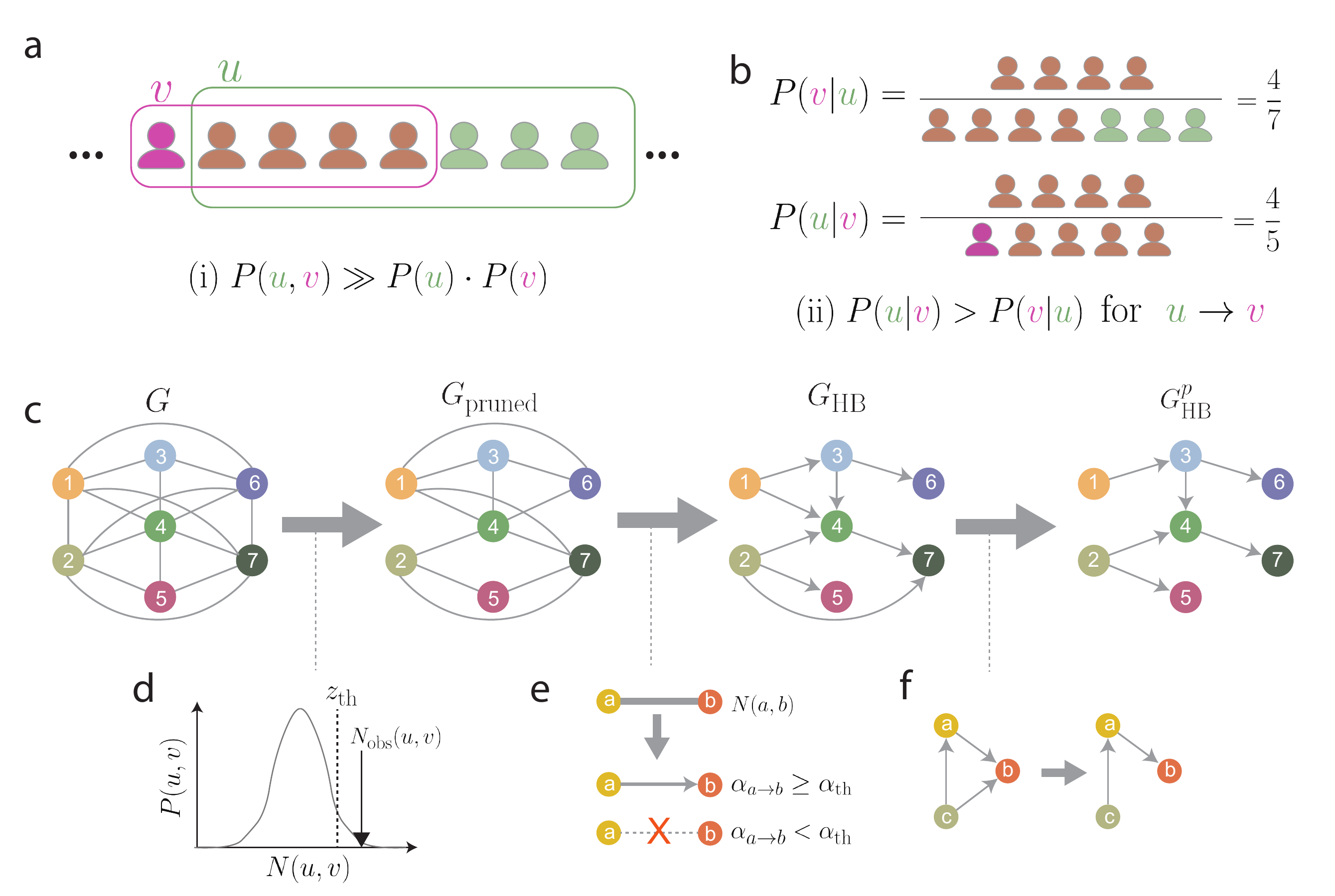}
\caption{
(a,b) An illustration of our approach. A user-tag bipartite network is displayed as a set cover, where two tags $u$ and $v$ have a hierarchical relationship $u \rightarrow v$.
As we expected, $u$ and $v$ co-occurs more frequently than expected, and there is an asymmetry in the conditional probability: $P(u|v) > P(v|u)$.
(c) The illustration of our method. First, we measure the $z$-score of the observed co-occurrence between tags $N_{\rm obs}(u,v)$ and remove the edges if the $z$-score does not exceed a threshold $z_{\rm th}$, obtaining $G_\text{pruned}$.
(d) We then calculate the hierarchical strength $\alpha_{u,v}$ for the remaining edges and threshold with $\alpha_{\rm th}$.
(e) Optionally, we make the backbone parsimonious by eliminating redundant edges.
}\label{fig:simple_model_desc}
\end{figure}

Let us consider a bipartite network $B$ in Fig.~\ref{fig:simple_model_desc}a, which consists of two sets of nodes: $\mathcal{O}$ (objects) and $\mathcal{T}$ (tags).
Our objective is to extract the hierarchical relationships between tags $\mathcal{T}$, although the approach is in principle applicable to either set of nodes.
Imagine a pair of tag nodes $u \in \mathcal{T}$ and $v \in \mathcal{T}$ that possesses an implicit hierarchical relationship, where $v$ can be considered as a sub-category of $u$, denoted as $u \rightarrow v$.
Then, we expect that, (i) $N(u,v)$, the number of nodes in $\mathcal{O}$ that are connected to both $u$ and $v$ in the bipartite network, is significantly larger than the expected co-occurrence $\mathbb{E} \left(N(u,v) \right)$ computed from an ensemble of degree-preserved random bipartite networks (co-occurrence), and
(ii) $P(u|v) \gg
P(v|u)$, where  $P(u|v) = \frac{N(u,v)}{N(v)}$, which also implies that $N(u) > N(v)$ (asymmetric conditional probability).
Figure~\ref{fig:simple_model_desc}a,b illustrate a simple example.

Our method begins with a one-mode projection of bipartite network $B$ onto tag nodes, which produces $G$, a unipartite network of tags (see Fig.~\ref{fig:simple_model_desc}c).
From $G$, we filter out the insignificant connections between tag nodes (condition (i)) by using $z$-score of observed co-occurrence $N(u,v)$ in $B$ from the probability distribution of expected co-occurrences~\cite{Tibey2013}.
Assuming the bipartite configuration model where degrees are preserved, the probability distribution of expected co-occurrence follows a hypergeometric distribution, yielding
\begin{eqnarray}
m &=& \frac{N(u) \cdot N(v) }{|\mathcal{O}|}, \nonumber
\\
\sigma^2 &=& \frac{N(u) \cdot N(v)}{|\mathcal{O}|}
\frac{|\mathcal{O}| - N(u)}{|\mathcal{O}|}
\frac{|\mathcal{O}| - N(v)}{|\mathcal{O}| -1},
\label{eq:hypergeometric}
\end{eqnarray}
where $m$ is the expected number of co-occurrences, $\sigma$ is its standard deviation, and $|\mathcal{O}|$ is the number of object nodes in $B$.
We remove connections between $u$ and $v$ in $G$
if
\begin{equation}
z_{u,v}  = \frac{N(u,v) - m}{\sigma} < z_{\rm th},
\end{equation}
where $z_\text{th}$ is a threshold parameter.

From the resulting $G_{\rm pruned}$, we estimate the hierarchical strength and direction between each pair of nodes by using the following functional form:
\begin{equation} \label{eq:hierarchy_strength_general_form}
\alpha_{u \rightarrow v} = f(k_u, k_v) \left( P(u|v) - P(v|u) \right),
\end{equation}
where $P(u|v)$ and $P(v|u)$ are the conditional probabilities defined above and $f(k_u, k_v)$ is a function that takes into account the importance of $u$ and $v$ in terms of their degrees $k_u$ and $k_v$ in $G_{\rm pruned}$.
Here, we adopt $f(k_u, k_v) = \frac{\min(k_u, k_v)}{k_{\max}}$ (where $k_{\max}$ is the maximum degree in $G_{\rm pruned}$) to preferentially extract the large-degree nodes, arriving at the following formula:
\begin{equation}\label{eq:hierarchy_strength}
\alpha_{u \rightarrow v} = \frac{\min(k_u, k_v)}{k_{\max}} \left( \frac{N(u, v) }{N(v)} -\frac{N(u, v) }{N(u)}  \right),
\end{equation}
where $\alpha_{u \rightarrow v} > 0$ indicates $u \rightarrow v$.
We interpret the magnitude of $\alpha_{u \rightarrow v}$ as the strength of the hierarchical relationship.
The hierarchical backbone $G_{\rm HB}$ of tag nodes in Fig.~\ref{fig:simple_model_desc} is constructed by collecting edges with $\alpha \geq \alpha_{\rm th}$.
$G_{\rm HB}$ is always a directed acyclic graph (DAG) because a node's position in the hierarchy is determined by its degree $N(u)$.
As we include more edges in the backbone (a smaller value of $\alpha_{\rm th}$), the extracted backbone would contain more redundant edges.
For instance, given the true hierarchical relationship $u \rightarrow v \rightarrow w$, it is likely to see not only $u \rightarrow v$ and $v \rightarrow w$, but also $u \rightarrow w$.
Hence, we add an optional procedure of making the hierarchical backbone $G_{\rm HB}$ parsimonious, removing the redundant edges that connect two nodes that are indirectly connected.

\subsection{Evaluation}

We first evaluate the performance of our method using semi-synthetic benchmarks, adopting the strategy from a previous study~\cite{Palla2015}.
To create benchmark networks, we use the currently documented hierarchical relationships from the Gene Ontology (GO)~\cite{Ashburner2000, Consortium2019} where gene products ($\mathcal{O}$) are tagged with GO terms ($\mathcal{T}$).
GO terms are classified into three domains: ``Biological Processes (BP)'', ``Cellular Components (CC)'', and ``Molecular Functions (MF)''.
Each GO term belongs to one of the domains and connected to its parents through three possible types of relations: ``is a'', ``part of'', and ``regulates''.
The reference hierarchical network ${\bar H}$ is extracted by gathering the GO terms belonging to the MF domain and all the edges that belong to either ``is a'' or ``part of'' types.
${\bar H}$, which is a DAG, consists of 11,078 GO terms and 13,773 directed edges between them.
From this reference hierarchy network, we generate an ensemble of synthetic bipartite networks based on the assumption that the related GO terms in ${\bar H}$ would appear more frequently together in the same gene product~\cite{Tibey2013}.
For each generated bipartite network, we first prepare $N$ ($N$ varies from $10,000$ to $500,000$ in our experiments) \emph{virtual} gene products.
For each gene product, we set the number of GO terms to be assigned by choosing a random number uniformly between 3 and 5, and randomly choose a `reference' GO term for the gene product.
To assign the remaining GO terms, we either randomly pick a GO term from the all possible set of GO terms (with probability $1-P_{RW}$) or perform a random walk from the reference GO term (with probability $P_{RW}$).
Each random walker, starting from the reference GO term, walks $s$ steps on ${\bar H}$, ignoring the directions of the edges, where $s$ is uniformly chosen between 1 and 3.
We attach the GO term where the random walker arrives at the gene product.
By repeating the same process, we obtain a bipartite network between the virtual gene products and GO terms.
We make 20 ensembles for each $N$ and evaluate our method with existing baselines~\cite{Tibey2013} across $N$ and $P_{RW}$ parameters.

The baseline algorithm A~\cite{Tibey2013} utilizes a projected GO term network with weights given by co-occurrence.
It detects hierarchical relations by computing $z$-score for co-occurrence between two GO terms and merge fragmented DAGs using the order of entropy of incoming weights for each local root.
The algorithm B~\cite{Tibey2013} employs eigenvector centrality on the weighted GO term network used in algorithm A.
The predicted edges of these baseline algorithms and ours are compared to the reference GO term hierarchy.
We use two types of comparisons: (i) \emph{edge-based} comparison that directly compares whether each predicted edge exists in the reference GO DAG, and (ii) \emph{path-based} comparison that checks whether each node pairs that are predicted to be connected are connected in the DAG through either a direct edge or a path.

\begin{figure}[h]
\includegraphics[width=\columnwidth]{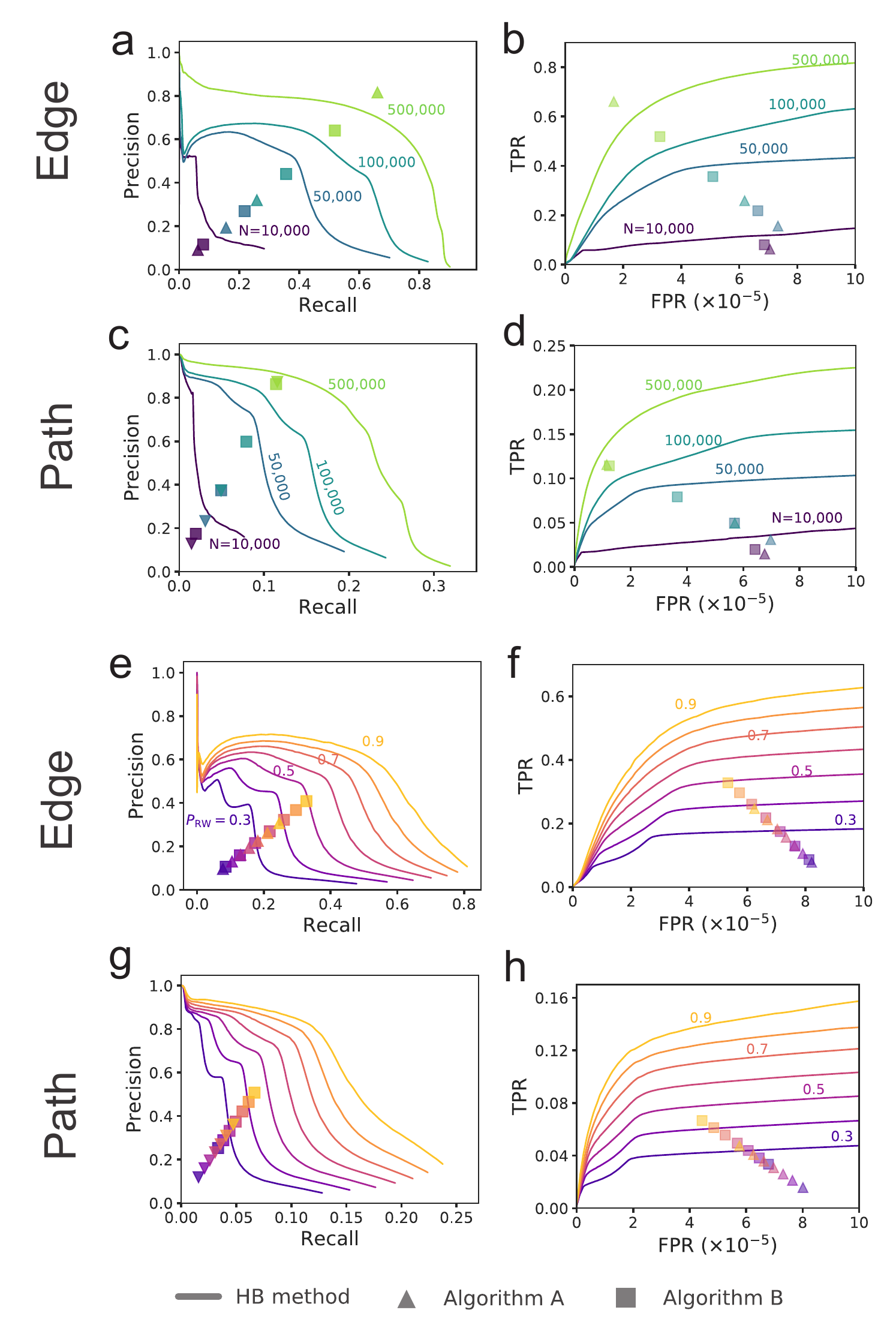}
\caption{
Quantitative evaluation with semi-synthetic benchmarks, where we vary the sample size $N$ and the random walk probability $P_\text{RW}$, demonstrates that our method (HB) outperforms the baselines in almost all cases, particularly when the data is sparse (smaller $N$ and smaller $P_\text{RW}$).
``Edge'' refers to the edge-based comparison where we only consider the existence of a predicted edge as a true positive; ``Path'' refers to the path-based comparison where we also consider the existence of an indirect path between the nodes that are predicted to be hierarchically connected as true positives too.
We measure (a-c) true-positive rate (TPR) with respect to false-positive rate (FPR) and (d-f) precision-recall curves with varying sample size $N$ in (a-d) or $P_{\rm RW}$ in (e-h).
We generate baseline results with existing methods with tuned parameters used in the original paper~\cite{Tibey2013}, and thus all measures are shown as points.
}\label{fig:synthetic_results}
\end{figure}

The evaluation demonstrates that our method outperforms the baselines in almost all cases (see Fig.~\ref{fig:synthetic_results}).
The only exception where the baseline outperforms is when $N$ = 500,000, where most information was given.
When data is sparse, our method reliably outperforms the others.
The pattern is similar when we use the path-based evaluation, where our method outperforms others in all cases.

\begin{figure*}[ht]
\makebox[\textwidth][c]{\includegraphics[width=0.9\textwidth]{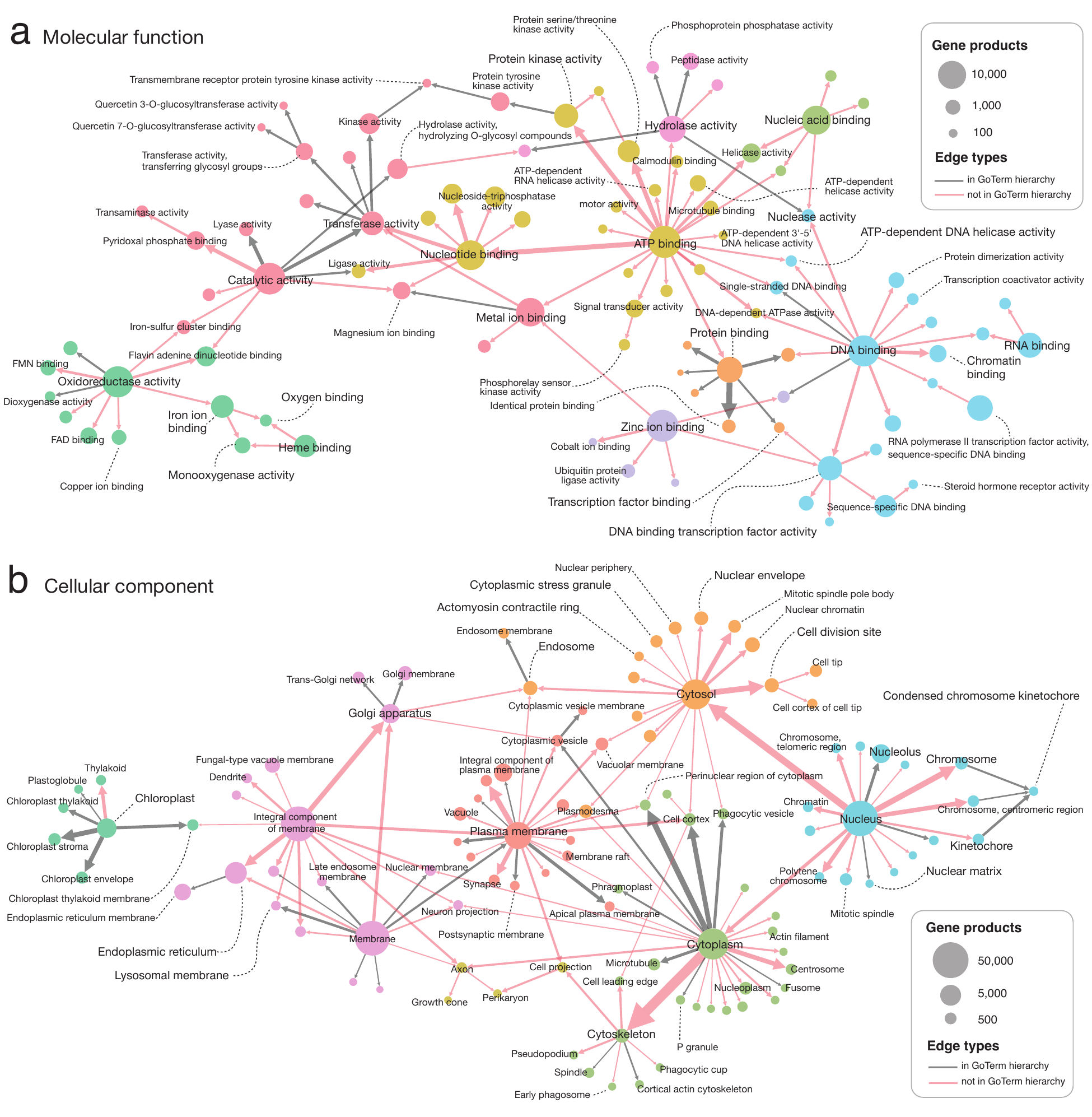}}
\caption{A hierarchical backbone of GO terms in the category of (a) molecular functions ($E=114$ with $\alpha > 0.03$) and (b) cellular components ($E=135$ with $\alpha > 0.064$).
We distinguish those that are also observed as paths in GO term ontology (roughly 25\% of links) and those that are not connected in the GO term ontology.
Node size corresponds to how many gene products are tagged with it.
Colors represent network communities~\cite{Blondel2008} found in the backbone.}\label{fig:hb_goterm}
\end{figure*}

As we establish our method's excellent performance in the semi-synthetic benchmarks, let us investigate the Gene Ontology (GO) hierarchy to examine the possibility of augmenting the existing ontology by identifying undocumented hierarchical relationships.
To do so, we merge the GO annotation datasets of 13 species (See Text S1 and Table S1), where each of the gene products is characterized by the GO terms assigned by experts, creating a bipartite network of gene products and GO terms.
From this bipartite network, we extract the hierarchical backbone for Molecular Function ($\alpha_{\rm th} = 0.03$; see Fig.~\ref{fig:hb_goterm}a) and Cellular Component ($\alpha_{\rm th} = 0.06$; see Fig.~\ref{fig:hb_goterm}b).

We show two types of edges: the black edges are those that exist both in the GO term hierarchy and in the edge set discovered from our method; the pink, augmented edges are those that are identified by our method but not currently in the Gene Ontology.
Our method organically discovers many biologically meaningful hierarchical relations---such as the edge between ATP-binding and Nucleotide-binding edge (See Fig.~\ref{fig:hb_goterm}(a)) and the nucleus-cytosol relationship (See Fig.~\ref{fig:hb_goterm}(b))---based on how gene products are annotated.

Taking a closer look into the backbone of Molecular Functions, which describes protein and enzyme functions, we find that the `Nucleotide binding', `DNA binding', and `ATP binding' lie at the center of the backbone and acquire many edges pointing to more specific binding-related functions.
ATP requirement for protein binding, protein kinase, and ATP-dependent enzymatic activities have been well documented~\cite{Kamerlin2013, Westheimer1987, Knowles1980} and are clearly represented in the relationship that extends out of the two main clusters.
The `DNA binding' connects to specific terms involving DNA, including `chromatin binding', `transcription control', `nuclease activity', and `transcription factor binding'.
In particular, note that `ATP-dependent DNA helicase activity'~\cite{Bochman2014} connects the `DNA binding' cluster to `ATP binding'.
The relationships between metabolic enzymatic activities are clearly present in the network too.
In the augmented hierarchy, `Catalytic activity' is connected to `Hydrolase activity', as well as `Oxidoreductase activity'  clusters.
Metal ions are commonly used to catalyze these enzymatic processes for metabolism and the maintenance of cellular redox status~\cite{Oteiza2012, Mohr2018}, which is captured by the proximity of the `Metal ion binding' cluster to these processes in the network.

The hierarchical backbone of the Cellular Component also reflects the cellular organizations well and captures nonexistent yet meaningful relationships between GO terms.
Our hierarchical backbone connects the `Nucleus' cluster (blue) to many components found within the nucleus but not documented in the Gene Ontology, including `Chromosome', `Nuclear matrix', `Kinetochores', and `Nucleolus'.
The `Cytosol' (orange) and the `Cytoplasm' (green) also acquire many edges in our algorithm;
Although one may consider that they describe the same cellular compartment, nuanced differences---which originates from the process of GO annotation---are captured in the hierarchical backbone.
The hierarchical backbone reveals that the `Cytosol' cluster primarily describes the parts relevant to the cell division process, ranging from the breakdown of the nuclear envelope, distribution of cytoplasmic cell granules, to the role of the actomyosin contractile ring during cytokinesis.
Meanwhile, the `Cytoplasm' cluster describes the cytoskeletal components involved in cell polarity, morphology, and movement.
The cell membranes are clustered as a group linking the cytosol and cytoplasm clusters, which highlights the membrane-based compartmentalization of many intracellular components in the cytoplasm/cytosol of the cell (Golgi apparatus, endosomes, endoplasmic reticulum, lysosomes, phagocytic vesicles, etc.).
As shown in these examples, the detected hierarchical backbone reveals meaningful ontological relationships between GO terms that are not present in the ontology but reflected in how the GO terms are used in the annotation practice.
We, therefore, argue that our method can potentially aid the process of building complex ontologies by suggesting ontological relationships from data.

\begin{figure*}[ht]
\includegraphics[width=\textwidth]{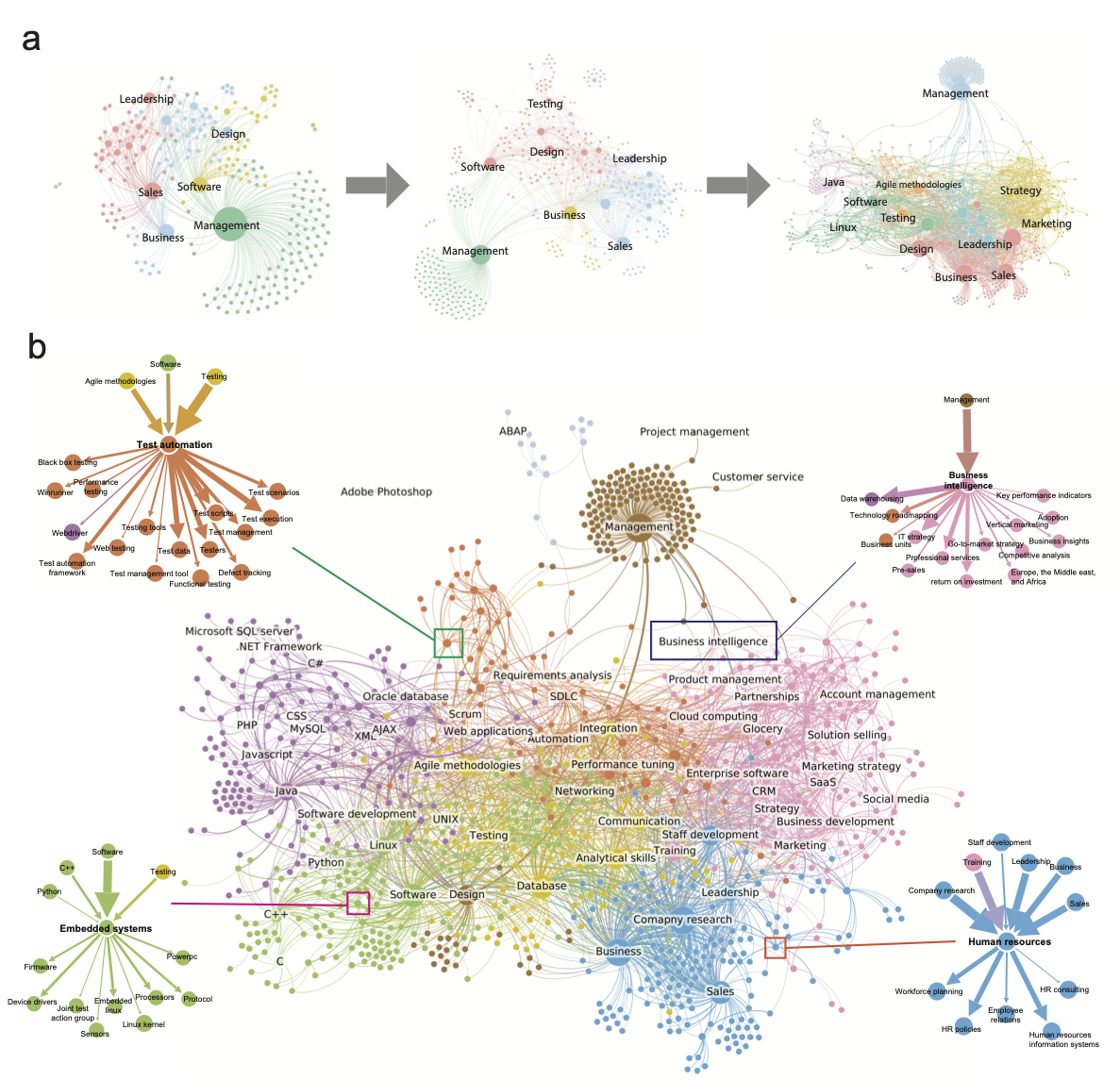}
\caption{ Hierarchical backbones of the software industry skills obtained from the LinkedIn's user-skill bipartite network.
(a) Hierarchical backbones with different thresholds ($\alpha_{\rm th} \approx 0.116, 0.092,$ and 0.067 from left to right with $z_{\rm th} = 20$), which determines the number of edges in a backbone ($E=500, 1000$ and $2000$, respectively).
As more edges are added, the backbone reveals more detailed structures around the hubs.
Colors represent the community membership found by the Louvain method within a backbone.
(b) The hierarchical backbone with $\alpha_{\rm th} =  0.05$ with $E=3346$.
The backbone reveals major skill clusters, which correspond to the common types of jobs in the software industry, including software engineering (left), general management (middle top), operational management (middle bottom), and market-based management (left).
We show the local neighborhoods of four skills: Test automation, Business intelligence, Human resources, and Embedded systems.}\label{fig:hb}
\end{figure*}

We then apply our method to a user-skill bipartite network dataset from LinkedIn, an online professional networking service~\cite{LinkedIn}.
From the LinkedIn users who identify themselves in the ``computer software'' industry on their profile and listed at least two skills, we sample about 340,000 users and their skills.
To remove skills that are too rare or erroneously typed, we dropped skills that appeared less than 100 times in our dataset, which left us with 6,736 skills.

We present several parsimonious hierarchical backbones across different $\alpha_{\rm th}$ in Fig.~\ref{fig:hb}(a) (we keep $z_{\rm th} = 20$ throughout the analysis).
At the early stage, our method identifies general, high-level skills such as `Management', `Sales', and `Software'; addition of more edges gradually reveals a complex, nuanced structure in the space of skills.
Delving into the hierarchical backbone shown in Fig.~\ref{fig:hb}(b) uncovers how the common skills in the software industry are organized.
For instance, it discovers the cluster of web-related skills (purple), which contains clusters of skills such as front-end (e.g., `Javascript', `PHP', `AJAX', and `CSS') and databases (e.g., `Oracle database' and `Microsoft SQL server').
Another major cluster (green) features skills related to the computer systems (i.e., `Linux', `maintenance', and `networking'), hardware (i.e., `embedded systems' and `x86') and analytical algorithms (i.e., `machine learning').
Management skills are clustered into three communities on the right side.
The blue community, located below in the backbone, exhibits traditional core skills on running a company, especially for human resource and finance, whereas the pink community is comprised of the skills for marketing, sales, service, and support, which follow traditional categories of functions in management~\cite{HBR2015}.
The brown community, located at the top, holds other general miscellaneous management skills.
Moreover, we found that the two clusters that connect the management
clusters and software development clusters are well-matched with the two new emerging functions in organizations~\cite{HBR2015}: (i) the yellow cluster bridges the software development with the service \& support, with the skills such as `Agile methodologies', `testing', and `communications', and (ii) the orange cluster connects different parts of the management sectors, which is often referred as ``customer success management'' in business literature~\cite{HBR2015}.
As illustrative examples, we sample four skills and their local hierarchical structures (ego-network) in Fig.~\ref{fig:hb}(b).
In sum, the hierarchical backbone of the software industry skills successfully captures detailed ontological relationships between skills purely from the bipartite network structure. In doing so, the backbone also reveals common roles in software firms, as well as confirming recent trends in the organizational structure.

\section{Conclusion}

In this study, we propose a method to extract a hierarchical backbone from a bipartite network by leveraging how implicit hierarchical relationships manifest in the bipartite network.
We show that our method consistently outperforms existing baseline methods by using semi-synthetic benchmarks that are derived from GO term hierarchy.
The application of our model to two real world datasets---the Gene Ontology and a professional-skill network data from LinkedIn---demonstrates that the method can extract important hierarchical relationships from organically generated object-tag bipartite network datasets, demonstrating the promise for automatic creation or augmentation of ontologies in various domains.

There are several limitations of our study.
Our model requires the choice of two threshold parameters $z_{\rm th}$ and $\alpha_{\rm th}$, and it is not yet fully understood how optimal parameters can be determined.
The method depends on certain assumptions on how tags are connected to the objects and thus may not work properly when the assumptions are not valid.
Also, because our method relies on a few heuristic, a principled statistical formulation of the problem may produce better results.

Nevertheless, we argue that our method can be a practical tool for extracting a hierarchical or ontological structure that is implicitly encoded in the object-tag bipartite networks.
Our method can also be combined with other network analysis methods, such as community detection, as shown in the examples in our study.
As numerous datasets can be described as bipartite networks, and because one-mode projection of such structure destroys crucial information, we argue that it is critical to develop methods that fully leverages the implicit information in the bipartite networks.
Our method may open up new ways to understanding hierarchical organization of complex networks in many fields and can be leveraged to create or augment hierarchical ontologies.

\section*{Acknowledgement}
We thank Haixu Tang and Irene Newton for helpful discussion.
This material is based upon work in part supported by the Air Force Office of Scientific Research under award number FA9550-19-1-0391.
B.~J.~K. was supported by the National Research Foundation of Korea (NRF) grant funded by the Korea government (MSIT) Grant No. 2019R1A2C2089463. 

\section{Appendix}
\subsection{Gene Ontology and annotations}
\subsubsection{Gene Ontology} 

In this study, we use Gene Ontology database and annotation datasets for 13 species provided by Gene Ontology Consortium~\cite{go_site}.
Gene Ontology (GO) is a structured vocabulary to express specific functions of genes from different organisms~\cite{Consortium2008}. 
Gene Ontology Consortium provides up-to-date GO database which contains a set of GO terms and relations to other GO terms. 
GO terms are classified into three categories: biological processes (BP), cellular components (CC), and molecular functions (MF). Each GO terms should belong to one of three classes. 
The GO terms also have hierarchical relations to its parents with one of three types: ``is a'', ``part of'', and ``regulates''. 
We downloaded Gene Ontology database released at 06 Jan. 2018 and constructed a directed acyclic graph from hierarchical relations with two types: ``is a'' and ``part of''. 
There are 4,157 GO terms with 8,227 relations in cellular components, 11,154 GO terms with 14,031 relations in molecular functions, and 29,691 GO terms with 73,231 relations in biological processes.

\subsubsection{Annotation datasets} 

This consortium also provides annotation datasets of individual species, where gene products are annotated with GO terms to represent normal functions of each gene product. 
We selected 13 species in the GO consortium~\cite{go_site} which has more than 1,000 gene products. 
For analysis, we excluded database for multi-species and mammals. 
The table~\ref{table_species} shows 13 species with size of gene products, annotated GO terms, and the release date. 
Annotation datasets of 13 species are provided in the archived GO consortium website now~\cite{go_archive}. 

\begin{table*}[]
\begin{center}
\begin{tabular}{c c c c c}\toprule 
& Species &  Gene products & Annotations & Date  \\  \colrule
1 & Arabidopsis thaliana & 25,726 & 49,876 & 12/31/2017  \\ 
2 & Aspergillus nidulans & 91,111 & 126,952 &  12/31/2017 \\ 
3 & Caenorhabditis elegans & 11,436 & 27,397 & 06/16/2017 \\ 
4 & Candida albicans & 42,007 & 124,841 & 09/24/2017 \\ 
5 & Danio rerio & 17,878 & 54,357 & 12/15/2017 \\ 
6 & Dictyostelium discoideum & 7,117 & 18,174 & 09/15/2017 \\ 
7 & Drosophila melanogaster & 13,318 & 43,704 & 09/22/2017 \\ 
8 & Escherichia coli & 2,647 & 5,741 & 12/08/2017 \\ 
9 & Oryza sativa & 1,752 & 2,547 & 10/28/2016 \\ 
10 & Plasmodium falciparum & 1,292 & 1,704 & 12/08/2017 \\ 
11 & Saccharomyces cerevisiae & 6,445 & 27,345 & 08/18/2017 \\ 
12 & Schizosaccharomyces pombe & 5,395 & 14,562 & 12/15/2017 \\ 
13 & Trypanosoma brucei & 2,523 & 3,092 & 12/01/2017 \\ \botrule
\end{tabular}
\end{center}
\caption{13 species for the analysis of GO terms of hierarchical backbones. The table shows a name, the number of gene products, the number of annotations, and the release date of each species.}\label{table_species}
\end{table*}

\bibliographystyle{unsrt}

\bibliography{main.bib}

\begin{thebibliography}{10}

\bibitem{Simon1962}
H.~Simon.
\newblock {The Architecture of Complexity}.
\newblock {\em Proceedings of the American Philosophical Society},
  106(6):467--482, 1962.

\bibitem{Clauset2008}
A.~Clauset, C.~Moore, and M.~E.~J. Newman.
\newblock {Hierarchical structure and the prediction of missing links in
  networks}.
\newblock {\em Nature}, 453(7191):98--101, 2008.

\bibitem{Blondel2008}
V.~D. Blondel, J.~L. Guillaume, R.~Lambiotte, and E.~Lefebvre.
\newblock {Fast unfolding of communities in large networks}.
\newblock {\em Journal of Statistical Mechanics: Theory and Experiment},
  2008(10):P10008, oct 2008.

\bibitem{yyAhn2010}
Y.-Y. Ahn, J.~P. Bagrow, and S.~Lehmann.
\newblock {Link communities reveal multiscale complexity in networks.}
\newblock {\em Nature}, 466(7307):761--764, 2010.

\bibitem{Sales-Pardo-Amaral}
M.~Sales-Pardo, R.~Guimer{\`a}, A.~A. Moreira, , and L.~A.~N. Amaral.
\newblock Extracting the hierarchical organization of complex systems.
\newblock {\em Proceedings of the National Academy of Sciences},
  104(39):15224--15229, 2007.

\bibitem{Hidalgo2007}
C.~A. Hidalgo, B.~Klinger, A.-L. Barab{\'{a}}si, and R.~Hausmann.
\newblock {The Product Space Conditions the Development of Nations}.
\newblock {\em Science}, 317(5837):482--487, jul 2007.

\bibitem{Clauset_hiring}
A.~Clauset, S.~Arbesman, and D.~B. Larremore.
\newblock {Systematic inequality and hierarchy in faculty hiring networks}.
\newblock {\em Science Advances}, 1(1):1--7, 2015.

\bibitem{Serrano2009}
M.~\'{A}. Serrano, M.~Bogu{\~{n}}{\'{a}}, and A.~Vespignani.
\newblock {Extracting the multiscale backbone of complex weighted networks.}
\newblock {\em Proceedings of the National Academy of Sciences of the United
  States of America}, 106(16):6483--8, 2009.

\bibitem{Radicchi2011}
F.~Radicchi, J.~J. Ramasco, and S.~Fortunato.
\newblock {Information filtering in complex weighted networks}.
\newblock {\em Physical Review E}, 83(4), 2011.

\bibitem{Grady2012}
D.~Grady, C.~Thiemann, and D.~Brockmann.
\newblock {Robust classification of salient links in complex networks}.
\newblock {\em Nature Communications}, 3(May):864, 2012.

\bibitem{Onnela2007}
J.-P. Onnela, J.~Saram{\"{a}}ki, J.~Hyv{\"{o}}nen, G.~Szab{\'{o}}, D.~Lazer,
  K.~Kaski, J.~Kert{\'{e}}sz, and A.-L. Barab{\'{a}}si.
\newblock {Structure and tie strengths in mobile communication networks.}
\newblock {\em Proceedings of the National Academy of Sciences of the United
  States of America}, 104(18):7332--7336, 2007.

\bibitem{Newman2003}
M.~E.~J. Newman and J.~Park.
\newblock {Why social networks are different from other types of networks}.
\newblock {\em Physical Review E}, 68:36122, 2003.

\bibitem{Balogh2019}
S.~G. Balogh, D.~Zagyva, P.~Pollner, and G.~Palla.
\newblock {Time evolution of the hierarchical networks between PubMed MeSH
  terms}.
\newblock {\em PLOS ONE}, 14(8):e0220648, 2019.

\bibitem{Dunne2002}
J.~A. Dunne, R.~J. Williams, and N.~D. Martinez.
\newblock {Food-web structure and network theory: The role of connectance and
  size.}
\newblock {\em Proceedings of the National Academy of Sciences of the United
  States of America}, 99(20):12917--22, 2002.

\bibitem{Goh2007}
K.-I. Goh, M.~E. Cusick, D.~Valle, B.~Childs, M.~Vidal, and A.-L.
  Barab{\'{a}}si.
\newblock {The human disease network}.
\newblock {\em Proceedings of the National Academy of Sciences of the United
  States of America}, 104(21):8685--8690, 2007.

\bibitem{Barabasi2016}
A.-L. Barab{\'{a}}si.
\newblock {\em {Network Science}}.
\newblock Cambridge University Press, 2016.

\bibitem{Barabasi1999}
A.-L. Barab{\'{a}}si and R.~Albert.
\newblock {Emergence of Scaling in Random Networks}.
\newblock {\em Science}, 286(5439):509--512, 1999.

\bibitem{Amaral2000}
L.~A.~N. Amaral, A.~Scala, M.~Barthelemy, and H.~E. Stanley.
\newblock {Classes of small-world networks}.
\newblock {\em Proceedings of the National Academy of Sciences},
  97(21):11149--11152, 2000.

\bibitem{Newman2001}
M.~E.~J. Newman.
\newblock {The structure of scientific collaboration networks}.
\newblock {\em Proceedings of the National Academy of Sciences},
  98(2):404--409, 2001.

\bibitem{Newman2002}
M.~E.~J. Newman, D.~J. Watts, and S.~H. Strogatz.
\newblock {Random graph models of social networks}.
\newblock {\em Proceedings of the National Academy of Sciences of the United
  States of America}, 99:2566--2572, 2002.

\bibitem{Zhou2007}
T.~Zhou, J.~Ren, M.~Medo, and Y.-C. Zhang.
\newblock {Bipartite network projection and personal recommendation}.
\newblock {\em Physical Review E}, 76(4):1--7, 2007.

\bibitem{Lehmann2008}
S.~Lehmann, M.~Schwartz, and L.~K. Hansen.
\newblock {Biclique communities}.
\newblock {\em Physical Review E}, 78(1):016108, jul 2008.

\bibitem{Larremore2014}
D.~B. Larremore, A.~Clauset, and A.~Z. Jacobs.
\newblock {Efficiently inferring community structure in bipartite networks}.
\newblock {\em Physical Review E}, 90(1):012805, 2014.

\bibitem{Bishop_book}
C.~M. Bishop.
\newblock {\em Pattern Recognition and Machine Learning (Information Science
  and Statistics)}.
\newblock Springer-Verlag New York, Inc., Secaucus, NJ, USA, 2006.

\bibitem{Heymann2006}
P.~Heymann and H.~Garcia-Molina.
\newblock {Collaborative Creation of Communal Hierarchical Taxonomies in Social
  Tagging Systems}.
\newblock {\em Stanford InfoLab Technical Report}, 10:1--5, 2006.

\bibitem{Schmitz2006}
P.~Schmitz.
\newblock {Inducing Ontology from Flickr Tags}.
\newblock {\em Collaborative Web Tagging Workshop}, pages 210--214, 2006.

\bibitem{Tibey2013}
G.~Tib{\'{e}}ly, P.~Pollner, T.~Vicsek, and G.~Palla.
\newblock {Extracting tag hierarchies}.
\newblock {\em PLoS ONE}, 8(12), 2013.

\bibitem{Palla2015}
G.~Palla, G.~Tib{\'{e}}ly, E.~Mones, P.~Pollner, and T.~Vicsek.
\newblock {Hierarchical networks of scientific journals}.
\newblock {\em Palgrave Communications}, 1:15016, 2015.

\bibitem{Ashburner2000}
M.~Ashburner, C.~A. Ball, J.~A. Blake, D.~Botstein, H.~Butler, J.~M. Cherry,
  A.~P. Davis, K.~Dolinski, S.~S. Dwight, J.~T. Eppig, M.~A. Harris, D.~P.
  Hill, L.~Issel-Tarver, A.~Kasarskis, S.~Lewis, J.~C. Matese, J.~E.
  Richardson, M.~Ringwald, G.~M. Rubin, and G.~Sherlock.
\newblock {Gene Ontology: tool for the unification of biology}.
\newblock {\em Nature Genetics}, 25(1):25--29, 2000.

\bibitem{Consortium2019}
The Gene~Ontology Consortium.
\newblock {The Gene Ontology Resource: 20 years and still GOing strong}.
\newblock {\em Nucleic Acids Research}, 47(D1):D330--D338, jan 2019.

\bibitem{Kamerlin2013}
Shina C.~L. Kamerlin, P.~K. Sharma, R.~B. Prasad, and A.~Warshel.
\newblock {Why nature really chose phosphate}.
\newblock {\em Quarterly Reviews of Biophysics}, 46(1):1--132, 2013.

\bibitem{Westheimer1987}
F.~H. Westheimer.
\newblock {Why nature chose phosphates}.
\newblock {\em Science}, 235(4793):1173--1178, 1987.

\bibitem{Knowles1980}
J.~R. Knowles.
\newblock {Enzyme-Catalyzed Phosphoryl Transfer Reactions}.
\newblock {\em Annual Review of Biochemistry}, 49(1):877--919, 1980.

\bibitem{Bochman2014}
M.~L. Bochman.
\newblock {Roles of DNA helicases in the maintenance of genome integrity}.
\newblock {\em Molecular {\&} Cellular Oncology}, 1(3):e963429, 2014.

\bibitem{Oteiza2012}
P.~I. Oteiza.
\newblock {Zinc and the modulation of redox homeostasis}.
\newblock {\em Free Radical Biology and Medicine}, 53(9):1748--1759, 2012.

\bibitem{Mohr2018}
S.~E. Mohr and D.~W. Killilea.
\newblock {Editorial: Metal Biology Takes Flight: The Study of Metal
  Homeostasis and Detoxification in Insects}.
\newblock {\em Frontiers in Genetics}, 9:221, 2018.

\bibitem{LinkedIn}
Linked{I}n, \url{http://www.linkedin.com}.

\bibitem{HBR2015}
M.~E. Porter and J.~E. Heppelmann.
\newblock {How smart, connected products are transforming companies}.
\newblock {\em Harvard Business Review}, 2015, 2015.

\bibitem{go_site}
Gene~Ontology Consortium.
\newblock Gene ontology consortium, 2018.

\bibitem{Consortium2008}
{The Gene Ontology Consortium}.
\newblock {The Gene Ontology project in 2008}.
\newblock {\em Nucleic Acids Research}, 36:D440--D444, 2008.

\bibitem{go_archive}
{The archive of the Gene Ontology Consortium}.
\newblock
  \url{http://www-legacy.geneontology.org/GO.downloads.annotations.shtml}.

\end{thebibliography}

\end{document}